\documentclass[12pt,preprint]{aastex}

\begin{document}

\title{THE COSMIC RADIO AND INFRARED BACKGROUNDS CONNECTION}

\author{Eli Dwek}
\affil{Laboratory for Astronomy and Solar Physics, NASA Goddard Space Flight Center,
Greenbelt, MD 20771, e-mail: eli.dwek@gsfc.nasa.gov}

\and

\author{Michael K. Barker}
\affil{Astronomy Department, University of Florida, Gainesville, FL 32611--2055, e-mail mbarker@astro.ufl.edu}

\begin{abstract}
We use the radio--infrared (IR) flux correlation between star--forming galaxies in the local universe to examine the connection between their cumulative contributions to the cosmic infrared and radio backgrounds. The general expression relating the intensities of the two backgrounds is complicated, and depends on details of the evolution of the galaxies' IR luminosity function with redshift. However, in the specific case when the radio--IR flux correlation is linear, the relation between the intensity of the IR background and the brightness temperature of the radio background reduces to a simple analytical expression which at 178~MHz is: $I_{\rm CIB}$(nW m$^{-2}$ sr$^{-1}$) = 2.7$\times \ T_{\rm crb}$(K), where the numerical coefficient was calculated for a radio spectral index of 0.7. This relation is insensitive to the star formation history of the galaxies that produce the cosmic IR background (CIB). We use the observed CIB intensity to constrain the cosmic star formation history, and the relation between the CIB and the cosmic radio background (CRB) to constrain the relative contribution of star--forming galaxies to the CRB. Current limits on the CIB intensity predict a 178~MHz brightness temperature of $\sim$~18$\pm$9~K, about half of the 37$\pm$8~K inferred for an isotropic radio component. This suggests that star--forming galaxies and AGN contribute about equally to the CRB intensity at that frequency. 
\end{abstract}

\section{INTRODUCTION}
Star forming galaxies exhibit a remarkable correlation between their radio and infrared (IR) fluxes covering over four decades of IR flux intensities (Helou, Soifer, \& Rowan--Robinson 1985, and references therein). The correlation results from the fact that the radio and infrared fluxes are different manifestations of the physical processes associated with the lifecycle of the
same stellar objects (Helou \& Bicay 1993, Lisenfeld, V\"olk, \&
Xu 1996). Massive stars heat the dust in the interstellar medium,  form H~II regions, and after their explosive deaths
accelerate particles to cosmic ray energies, processes that,
respectively, give rise to the observed
galactic thermal infrared, radio thermal and synchrotron emission. The correlation may not hold for active galactic nuclei (AGN), in which the radio emission is not associated with the life cycle of massive stars.

If the radio--IR correlation for individual star forming galaxies in the local universe holds for all redshifts, then their cumulative contributions to the cosmic radio and infrared backgrounds should be related.  Haarsma \& Partridge (1998, hereafter HP98) used this correlation to estimate the intensity of the cosmic radio background (CRB) that can be attributed to star--forming galaxies.  At the time of their analysis only the CIB at wavelengths larger than $\sim$ 120 $\mu$m was definitively detected by the Diffuse Infrared Background Experiment (DIRBE) and Far Infrared Spectrophotometer (FIRAS) instruments on board the {\it Cosmic Background Explorer} ({\it COBE}) satellite (Hauser et al. 1998, Fixsen et al. 1998). For simplicity, HP98 assumed that all the sources giving rise to the CIB release their energy instantaneously at redshift $z$~=~1. With this simple assumption about the formation epoch of the CIB, most of its detected intensity is determined by the $\sim$ 80 $\mu$m spectral luminosity density of galaxies at the assumed redshift of its creation. Adopting a linear relation, $S_{\nu}(20\ cm) \propto S_{\nu}(80\ \mu$m), between the radio and IR fluxes from star--forming galaxies, HP98 derived their contribution to the 40 cm CRB intensity.  To compare this CRB estimate to the observed Bridle (1967) datum point at 178~MHz,  HP98 adopted a $\nu^{-0.7}$ power law spectrum for the radio sources. They derived a brightness temperature of $T _{\rm crb}$(178\ MHz) $\approx$~15~K, about half the value inferred from the observations. HP98 attributed the remaining CRB intensity to the contribution of AGN which do not contribute significantly to the CIB (Barger et al. 2001, Fadda et al. 2001). 

In this paper we extend the analysis of HP98 to the more general case in which the cosmic star formation rate (CSFR) of the CIB sources is a general function of redshift.  First, in \S2, we briefly review the various presentations of the radio--IR correlation and rederive this correlation in terms of the galaxies' 8 -- 1000 $\mu$m IR fluxes. Previous presentations expressed this correlation as a function of only the far--IR (FIR) fluxes, derived from the galaxies 60 and 100 $\mu$m detections by the {\it Infrared Astronomy Satellite} ({\it IRAS}). Such a correlation may systematically underestimate the radio flux from more luminous star forming galaxies in which an increasing fraction of the IR flux may be emitted in the shorter (12 and 25 $\mu$m) {\it IRAS} bands. We then explore the functional form of the radio--IR correlation. HP98 adopted a linear relation between the radio and IR fluxes, which may not be an accurate fit to the data. Significant deviations from linearity will require detailed knowledge of the evolution of the galaxies' IR luminosity function with redshift. However, we argue that the data are consistent with a linear relation over most of the range of the IR luminosities that contribute to the local IR luminosity density. In \S3 we first derive the expressions for the CIB and CRB intensities produced by star--forming galaxies for a general power--law correlation between their radio and IR fluxes. We then derive a simple analytic expression relating these background intensities when their radio--IR correlation is linear. In \S4 we apply these results to different cosmic star formation histories with their associated CIB intensity limits. A brief summary of our paper is presented in \S5.

\section{THE RADIO--INFRARED FLUX CORRELATION}

The radio--IR correlation can be written in a general form as:
%---------------------
\begin{equation}
P_{\nu}(\nu,\ {\rm L_{IR}})({\rm W\ Hz}^{-1}) = \kappa(\nu) \  \left( {{\rm L_{IR}} \over {\rm L}_{\odot}}\right)^{\beta(\nu)}
%P_{\nu}(\nu,\ L_{{\rm IR}})({\rm W\ Hz}^{-1}) = \kappa(\nu) \times L^{\beta(\nu)}_{\rm IR}({\rm L}_{\odot})
\end{equation}
%---------------------
where L$_{\rm{IR}}$ is the IR luminosity of the galaxy, $P_{\nu}(\nu)$ is the specific radio luminosity at the radio frequency $\nu$, and $\kappa$ and $\beta$ are coefficients that depend on the frequency at which the correlation is expressed.  The coefficient $\kappa$ is dimensional, and its value depends on the units used in the correlation. The presentation above is often referred to as the \{P$_{\nu}$--L$_{\rm IR}$\} presentation of the radio--IR correlation.  An alternative is the \{S$_{\nu}$--F$_{\rm IR}$\} presentation, in which the radio and IR outputs are expressed in units of W m$^{-2}$ Hz$^{-1}$ and W~m$^{-2}$, respectively.  

The two presentations are, of course, equivalent for individual galaxies. However, for determining $\beta$, the choice of presentation is important if $\beta$ is a priori known to be different from unity (Cox et al. 1988). If all galaxies lie on a line of slope $\beta$ = 1 in a log--log S$_{\nu}$ vs. F$_{IR}$ plot, then the conversion to a \{P$_{\nu}$--L$_{IR}$\} presentation will move them up or down along a diagonal line, leaving the overall slope of the correlation unchanged. However, if the slope of the correlation is different from unity in a \{S$_{\nu}$--F$_{IR}$\} presentation, then fainter galaxies, which tend to be more distant than brighter ones, may move systematically away or toward the $\beta$   = 1 slope line in a \{P$_{\nu}$--L$_{\rm IR}$\} presentation. The overal effect will be to flatten the slope if $\beta <$ 1, and to steepen it if $\beta >$ 1 in a transition from a \{S$_{\nu}$--F$_{\rm IR}$\} to a \{P$_{\nu}$--L$_{\rm IR}$\} presentation. 

Various investigators examined the radio--IR correlation using a variety of sample selection criteria. All use FIR fluxes measured  by the {\it IRAS} satellite and construct galaxy FIR luminosities (fluxes) from the 60 and 100 $\mu$m detections using the following relation (Sanders \& Mirabel 1996):
%---------------------
\begin{eqnarray}
F_{\rm {FIR}}({\rm\ W m}^{-2}) &  = & 1.26\times10^{-14}\ [2.58\ S_{\nu}(60 \mu m)+S_{\nu}(100 \mu m)] \\ \nonumber
{\rm L_{FIR}}({\rm L}_{\odot}) &  = & 1.26\times10^{12}\ [2.58\ {\rm L}_{\nu}(60 \mu m)+
{\rm L}_{\nu}(100 \mu m)] 
\end{eqnarray}
%--------------------
where $S_{\nu}$ and L$_{\nu}$ are flux and luminosity densities expressed in units of Jy ( = 10$^{-26}$ W m$^{-2}$ Hz$^{-1}$) and  L$_{\odot}$/Hz, respectively.

As we will show in \S3, the value of $\beta$ plays an important role in quantifying the CIB--CRB connection. A value of $\beta$ that is significantly different from unity will require detailed knowledge of the evolution of the IR luminosity function with redshift, whereas a value of unity will considerably simplify the relation between the two background emissions. We therefore briefly review previous determinations of $\beta$.

Initial investigations, using a relatively small sample of 91 radio--selected galaxies observed at $\nu$~=~4.8 GHz (de Jong et al. 1985) and 38 optically-- and radio-- selected galaxies observed at 1.4 GHz (Helou, Soifer, \& Rowan--Robinson 1985), found the radio--IR correlation to be linear.  Wunderlich \& Klein (1988) extended the analysis of de Jong et al. (1985) to a wider range of galaxy types, covering about 4 orders of magnitudes in FIR luminosities. They found that the correlation at $\nu$ = 4.8 GHz is approximately linear ($\beta = 0.99\pm0.38$) up to IR luminosities of  $9.1\times10^{9}$ L$_{\odot}$, with $\beta$ increasing to a value of $1.26\pm0.33$ at higher luminosities. However, we found that their sample of galaxies could also be fitted with a single power law with a slope of $\beta$ = 1.05$\pm$0.03. 

Devereux and Eales (1989) correlated the 1.49 GHz power with the FIR output from an optically selected sample of galaxies with luminosities between $\approx \ 10^8-10^{11}$ L$_{\odot}$. They found the slope of the correlation to be significantly larger than unity with a value of $\beta$ = 1.28.  A similar conclusion was reached by Cox et al. (1988) who studied the correlation for a flux limited sample of 74 radio galaxies at 151 MHz. They found $\beta$ = 1.32$\pm$0.06 in a \{S$_{\nu}$--F$_{IR}$\} presentation, and $\beta$ = 1.15$\pm$0.04 in a \{P$_{\nu}$--L$_{\rm IR}$\} presentation of the data, and adopted an average slope of $\beta$ = 1.23.  

Chi \& Wolfendale (1990) provided theoretical arguments why the slope of the radio--IR correlation should not be unity. They predicted  a break in the slope of the correlation at an IR luminosity above which most of the electrons producing the radio synchrotron emission remained trapped in the galaxy. For these galaxies the slope of the correlation should be unity, but for the smaller and less IR luminous galaxies they argued that $\beta$ should be larger than unity. They claimed to have found evidence for this effect in the data and reported values of $\beta$ = 1.37 for low luminosity galaxies, a trend exactly opposite to that found by Wunderlich \& Klein (1988). Our analysis of the Chi \& Wolfendale sample did not confirm either trend. We found that their data could be represented by a single slope  with a value of $\beta = 1.27 \pm 0.06$.

Condon, Anderson, \& Helou (1991; hereafter CAH91) reexamined the radio--IR correlation using the {\it IRAS} revised bright galaxies sample (BSG) from which spectroscopically identified AGN were subtracted, and which were detected with the VLA at 1.49 GHz. They  found the slope of the  correlation to be significantly greater than unity with a smaller value of $\beta = 1.11 \pm 0.02$ over the $\sim 10^9$ to 10$^{13}$ L$_{\odot}$ luminosity range.

For the purpose of the present analysis we reexamined the radio--IR correlation starting from the  same galaxy sample used by CAH91, the revised {\it IRAS} Bright Galaxy Sample, elimating galaxies that were not detected in all 4 {\it IRAS} bands, and removing AGN using a more recent catalog of spectroscopically confirmed AGN galaxies (Veron-Setty \& Veron 2001, catalog of Quasars and AGN). The total number of galaxies left in the sample is 222, compared to 258 galaxies used in the analysis of CAH91. We used the additional 12 and 25~$\mu$m {\it IRAS} bands to derive the total 8--1000~$\mu$m IR luminosity from the relation (Sanders \& Mirabel 1996):
%---------------------
\begin{equation}
{\rm L_{IR}}({\rm L}_{\odot}) =1.8\times10^{12}[13.48\ {\rm L}_{\nu}(12\ \mu m)+5.16\ {\rm L}_{\nu}(25\ \mu m)+2.58\ {\rm L}_{\nu}(60\ \mu m)+{\rm L}_{\nu}(100\ \mu m)]
\end{equation}
%--------------------

\begin{figure}
%\begin{center}
% \epsscale{0.5}
\plotone{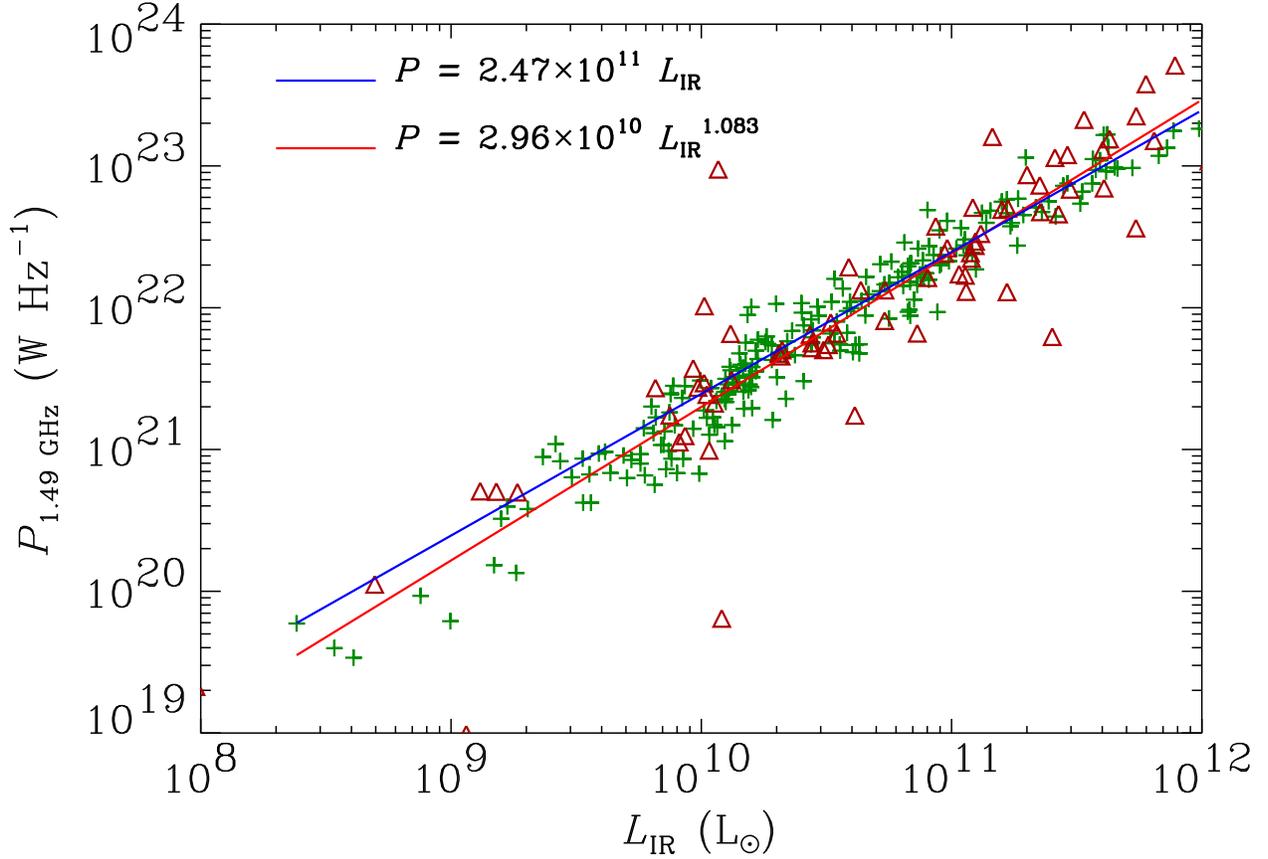}
\caption{The radio fluxes of galaxies in the {\it IRAS} BGS are plotted versus their 8--1000~$\mu$m luminosity determined from eq. (3). Star--forming galaxies are represented by crosses, and spectroscopically confirmed AGN by open triangles. The lines are fits to the radio--IR correlation between the star--forming galaxies only. The solid blue line represents the best power--law fit, and the red line the best linear fit to the data. The population of AGN generally follows the same correlation, but with a wider dispersion.}
\end{figure}

Figure 1 depicts the radio--IR correlation for a \{P$_{\nu}$--L$_{IR}$\} presentation of the data, and the best power-law fit to the data.  
We derive an almost identical slope $\beta$ for the same sample of galaxies as CAH91, with a smaller value for $\kappa$, reflecting the fact that the 8 -- 40 $\mu$m emission was not included in their galactic IR energy budget.  The values of the fit for radio--IR correlation are: \{$\kappa$,~$\beta$\} = \{2.96$\times10^{10}$,~$1.083 \pm0.02$\}. An almost identical slope was obtained for the  \{S$_{\nu}$--F$_{\rm IR}$\} presentation, suggesting that there were no significant flux related systematic errors in determining the distances of the galaxies in the sample. Also shown in Figure 1 is a forced linear fit to the data with \{$\kappa$, $\beta$\} = \{2.47$\times10^{11}$,~1.0\}. The linear fit is almost indistinguishable from the best fit for galaxies with luminosities above $\sim$ 10$^{10}$ L$_{\odot}$, which produce most of the IR luminosity density in the local and high--$z$ universe. The same linear fit will provide a similarly good representation of the radio--IR correlation  even if the AGN were included in the data. The AGN population  however shows more dispersion around the fit.
 
\section{THE CIB and CRB INTENSITIES PRODUCED BY STAR--FORMING GALAXIES}

The specific intensity $I^{\rm crb}_{\nu}(\nu_R)$ of the CRB at the observed radio frequency
$\nu_R$ is given  by the integral over sources (e.g. Peebles 1993):
%---------------------
\begin{equation}
I^{\rm crb}_{\nu}(\nu_R) = \left ({c\over 4 \pi}\right)\ \int_0^{\infty}
{\cal E}_{\nu}(\nu,z)\left|{dt\over dz}\right|dz
\end{equation}
%---------------------
\noindent
where  ${\cal E}_{\nu}(\nu, z$) is the specific radio luminosity per comoving volume element at redshift $z$,  $\nu = \nu_R(1+z)$ is the frequency in the rest frame of the luminous  objects and  $\left|dt/dz\right|$ is given by (Hogg 1999)
%---------------------
\begin{equation}
\left|dt/dz\right|^{-1}  =  H_0
(1+z)\left[(1+z)^2(1+\Omega_mz)-z(2+z)\Omega_{\Lambda}\right]^{1/2}
\end{equation}
%---------------------
where $H_0$ is the Hubble constant,  $h$ its value in units of 100~km~s$^{-1}$~Mpc$^{-1}$, $\Omega_m\equiv \rho_m/\rho_c$ is the present mass density of the universe normalized 
to the critical density $\rho_c = 1.88\times10^{-29}\ h^2$~g~cm$^{-3}$, and
$\Omega_{\Lambda} \equiv \Lambda/3H^2_0$ is the dimensionless cosmological constant. 

We will assume that the radio--IR correlation, expressed at a given rest frame frequency $\nu_0$, holds for all redshifts and that the radio spectrum of the individual galaxies follows a $\sim \nu^{-\alpha}$ power law with spectral index ${\alpha}$, so that 
$P_{\nu}(\nu,\ {\rm L_{IR}},\ z)=(\nu/\nu_0)^{-\alpha}P_{\nu}(\nu_0,\ {\rm L_{IR}},\ z)$.  The spectral luminosity density ${\cal E}_{\nu}$ can then be expressed as:
%---------------------
\begin{eqnarray}
{\cal E}_{\nu}(\nu,z) &  = &  \int_{\rm L_{min}}^{\rm L_{max}}  P_{\nu}(\nu, {\rm L_{IR}}, z)\ \Phi({\rm L_{IR}},\ z)\ {\rm dL_{IR}} \\ \nonumber
 &  = & \kappa_0\ \left({\nu \over \nu_0}\right)^{-\alpha}\ \int_{\rm L_{min}}^{\rm L_{max}}  \ {\rm L_{IR}}^{\beta_0}\ 
\Phi({\rm L_{IR}},\ z)\ {\rm dL_{IR}} 
\end{eqnarray}
%---------------------
where  $\Phi({\rm L_{IR}},\ z)  \equiv [{\rm d}\varphi({\rm L_{IR}},\ z) / {\rm dL_{IR}}]$ is the differential IR luminosity function, d$\varphi({\rm L_{IR}},\ z)$ is the comoving number density of galaxies with luminosities between L$_{\rm IR}$ and L$_{\rm IR}$ + dL$_{\rm IR}$ in the \{${\rm L_{min}},\ {\rm L_{max}}$\} luminosity interval, and $\kappa_0$, $\beta_0$ are the parameters of the radio--IR correlation at frequency $\nu_0$. 

The specific intensity of the CRB can now be written as:
%---------------------
\begin{equation}
 I^{\rm crb}_{\nu}(\nu_{_R}) =  \left ({c\over 4 \pi}\right)\ \left({\nu_{_R} \over \nu_0}\right)^{-\alpha}\  \int_0^{\infty} \ { f(z)\ {\cal L}_{\rm IR}(z)\ \over (1+z)^{\alpha}} \left|{dt\over dz}\right|\ dz
\end{equation}
%---------------------
where ${\cal L}_{\rm IR}$(z) is the comoving IR luminosity density at redshift $z$:
%---------------------
\begin{equation}
{\cal L}_{\rm IR}(z) \equiv  \int_{\rm L_{min}}^{\rm L_{max}}  {\rm L_{IR}}\  \phi({\rm L_{IR}},\ z)\ {\rm dL_{IR}}   
\end{equation}
%---------------------
and $f(z)$ is defined as:
%---------------------
\begin{equation}
f(z) \equiv  \kappa_0\ \int_{\rm L_{min}}^{\rm L_{max}} {\rm L_{IR}}^{\beta_0}   \phi({\rm L_{IR}},\ z)\ {\rm dL_{IR}} \over {\cal L}_{\rm IR}(z)  .  
\end{equation}
%---------------------
\noindent
The same star forming galaxies give rise to the CIB with an intensity $I_{\rm CIB}$ that is given by:
%---------------------
\begin{equation}
I_{\rm CIB}  = \left ({c\over 4 \pi}\right)\  \int_0^{\infty}\  { {\cal L}_{\rm IR}(z) \over 1+z} \left|{dt\over dz}\right|\ dz
\end{equation}
%---------------------

For a linear correlation, $\beta_0$ = 1, and $f(z) = \kappa_0$(Hz$^{-1}$). The spectral intensity of the CRB then becomes:
%---------------------
\begin{equation}
I^{\rm crb}_{\nu}(\nu_{_R})   =  \left ({c\over 4 \pi}\right)\ \left({\nu_{_R} \over \nu_0}\right)^{-\alpha}\  \kappa_0 \int_0^{\infty} \ { {\cal L}_{\rm IR}(z)\ \over (1+z)^{\alpha}} \left|{dt\over dz}\right|\ dz
\end{equation}
%---------------------
\noindent
In this case the relation between $I^{\rm crb}_{\nu_R}$ and $I_{\rm CIB}$ is considerably simplified and given by:
%---------------------
\begin{eqnarray}
I_{\rm CIB}  & = & \left ({\nu_R\over \nu_0}\right)^{\alpha}\ {g(\alpha)\over \kappa_0}\ I^{\rm crb}_{\nu}(\nu_{_R}) \nonumber \\
   & = & 1.18\times 10^8\ \left ({\nu_R\over \nu_0}\right)^{\alpha}\ {g(\alpha)\over \kappa_0}\ \nu^2_R(\rm {MHz})\ T_{\rm crb}(\nu_R)\ \ \ \ \rm{nW\ m}^{-2}\ \rm{sr}^{-1}
\end{eqnarray}
%---------------------
where
%---------------------
\begin{equation}
g(\alpha) \equiv  \left\{\int_0^{\infty}\  { {\cal L}_{\rm IR}(z) \over 1+z} \left|{dt\over dz}\right|\ dz \right\}  /
                                  \left\{\int_0^{\infty}\  { {\cal L}_{\rm IR}(z) \over (1+z)^{\alpha}} \left|{dt\over dz}\right|\ dz\right\} \ \ \ ,
\end{equation}
%---------------------
and $T_{\rm crb}(\nu_R)$ is the CRB brightness temperature at frequency $\nu_R$.
 
\section{THE CIB--CRB CONNECTION FOR VARIOUS COSMIC STAR FORMATION HISTORIES}
 
Several models have been recently developed for calculating the evolution of ${\cal L}_{\rm IR}(z)$ with redshift. Here we will focus on some select representative cases: Malkan \& Stecker (2001, baseline model), Chary \& Elbaz (2001, pure luminosity evolution model), and Xu et al. (2001, peak model). These models were developed to explain the limits and detections of the CIB spectrum and galaxy number counts obtained with the {\it IRAS}, the {\it Infrared Space Observatory} ({\it ISO}),  and the Submillimeter Common User Bolometric Array (SCUBA) instrument on the James Clerk Maxwell Telescope (JCMT) at various IR and submillimeter wavelengths. Figure 2 presents the evolution of  ${\cal L}_{\rm IR}(z)$ for these select models. Also plotted are observational estimates of the comoving cosmic star formation rate (CSFR) at different redshifts. Star formation rates were converted to IR luminosity densities using the relation (Kennicutt 1998): ${\cal L}_{\rm IR}$(L$_{\odot}$~Mpc$^{-3}$)~=~$6\times 10^{9}\times \rho_*$(M$_{\odot}$~yr$^{-1}$~Mpc$^{-3}$). References to the CSFR data can be found in the papers listed above. Also shown in the figure are approximate upper and lower limits to the IR luminosity density.   The top--hat function centered on redshift $z$~=~1 in the figure represents the instantaneous energy injection model used by Haarsma \& Partridge (1998) to calculate the CRB. The amplitude of the function was chosen so that it reproduced the nominal $\sim$ 5--1000 $\mu$m CIB intensity of 50 nW m$^{-2}$ sr$^{-1}$ (Hauser \& Dwek 2001). 

\begin{figure}
%\begin{center}
% \epsscale{0.5}
\plotone{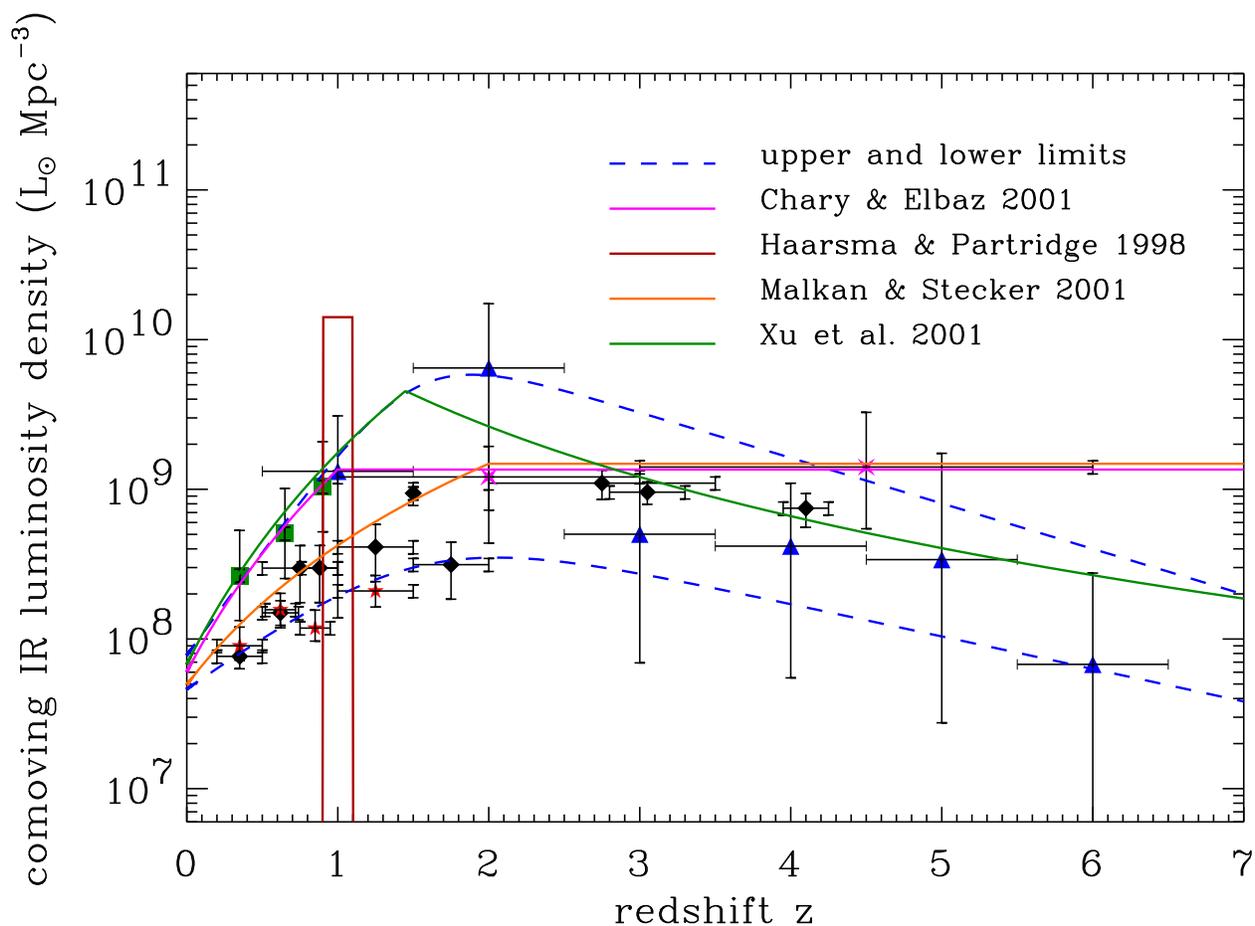}
\caption{The comoving IR luminosity density predicted by several galaxy number count models as a function of redshift. The data points represent optical and IR determinations of the comoving cosmic star formation rate as a function of redshift (references to the data can be found in the papers listed above). A scaling factor of $6\times10^{9}$ was used to convert the CSFR (in M$_{\odot}$ yr$^{-1}$)  to an IR luminosity density (in L$_{\odot}$  Mpc$^{-3}$). The dashed blue lines represent approximate upper and lower limits to the CSFR. }
\end{figure}

Figure 3 shows the value of $g(\alpha)$ as a function of $\alpha$ for the different star formation histories depicted in Figure 2. The function $g(\alpha)$ is insensitive to the cosmic star formation history, a direct consequence of the fact that the integrands in eq. (13) differ only by a factor of $(1+z)^{\alpha-1}$. For $\alpha\ \approx$ 0.6--1.2, the most probable range of values for the radio spectral index, $g(\alpha)$ is well approximated by:
%---------------------
\begin{equation}
g(\alpha)  =  \alpha^{0.688} \ \ \ \ \  \rm{ for}\ \ \ \ \alpha \approx 0.6-1.2
\end{equation}
%---------------------
to an accuracy better than 2\%.

\begin{figure}
%\begin{center}
% \epsscale{0.5}
\plotone{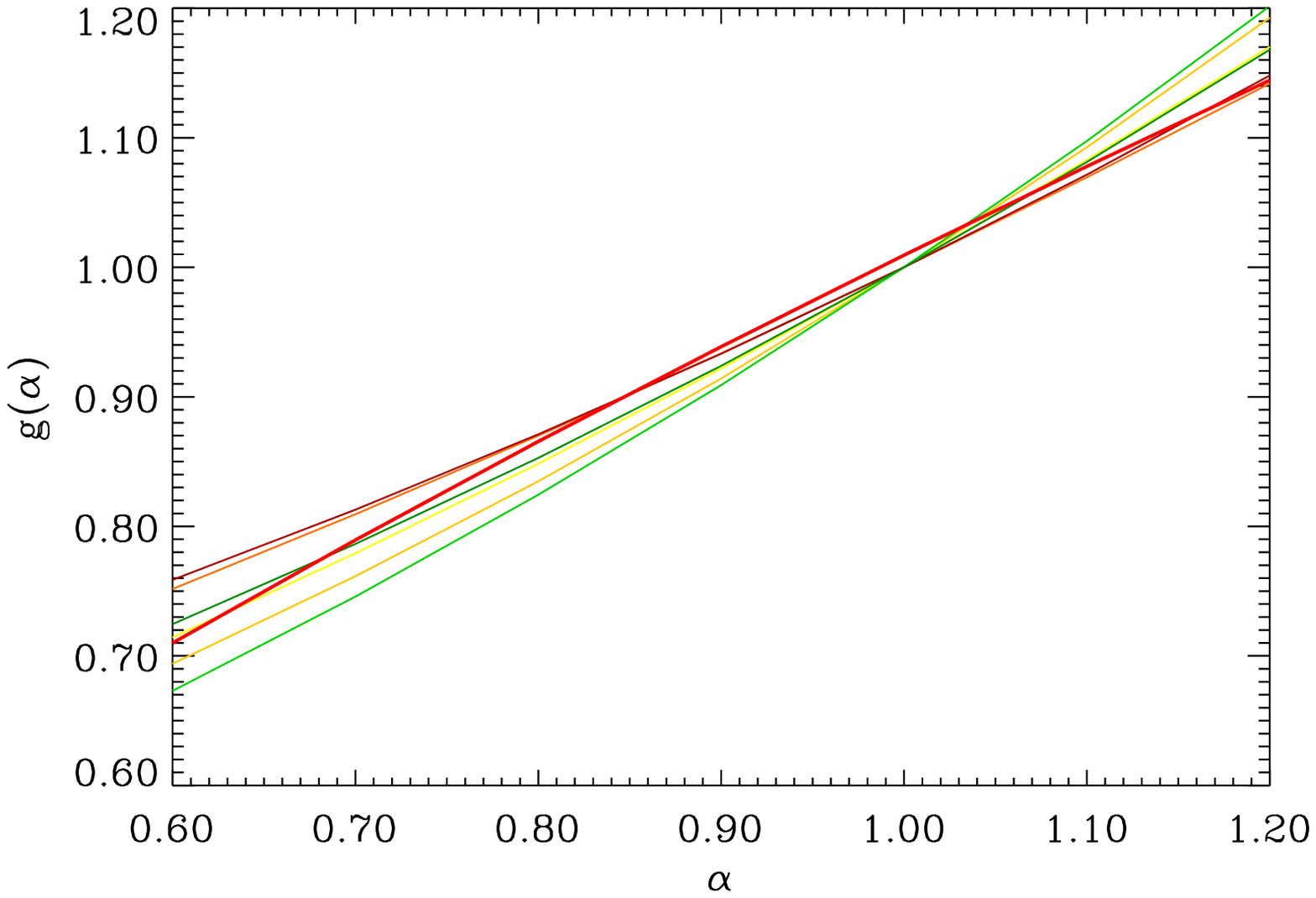}
\caption{The function $g(\alpha)$, defined by eq. (13) is plotted against $\alpha$ for the various CSFR depicted in Figure 2. The thick red line represents a $\alpha^{0.688}$ power--law fit to the function for $\alpha$ values between 0.6 and 1.2. }
\end{figure}

Using eqs. (12), (14) and the linear radio--IR correlation at $\nu_0$ = 1.49 GHz: $P_{\nu}(\rm{W\ Hz}^{-1})~=~2.47\times 10^{11}\ L_{\rm IR}$(L$_{\odot}$), we can express the CIB intensity in terms of the radio brightness temperature at frequency $\nu_R$~=~178~MHz as:
%---------------------
\begin{eqnarray}
I_{\rm CIB}(\rm {nW m}^{-2} \rm{ sr}^{-1})  & = &  {\cal A}(\alpha)\  T_{\rm crb}  \nonumber \\
\rm{where}\ \ \ {\cal A}(\alpha) & = &  \{3.0,\ 2.7,\ 2.4,\ 2.1\} \ \ \ \ \  \rm{ for}\ \ \ \ \alpha = \{0.6,\ 0.7,\ 0.8,\ 0.9\}
\end{eqnarray}
%---------------------

Table 1 lists the values of the CIB intensity and the 178~MHz brightness temperature obtained from the respective use of eqs. (10) and (11) for the different star formation histories and radio spectral index $\alpha$. The entries in the table satisfy the analytical approximation for the relation between $I_{\rm CIB}$, $T_{\rm crb} $, and $\alpha$ to an accuracy of a few percent. For comparison we also listed the observed limits and detections of the CIB (Hauser \& Dwek 2001) and the 178~MHz CRB brightness temperature inferred from the Bridle data (Bridle 1967) for the different values of $\alpha$. Also shown in the table are the observed $T_{\rm crb}/I_{\rm CIB}$ ratios and those predicted by the CIB-CRB correlation [eq. (15)].

It is interesting to compare our model predictions with the result obtained by Haarsma \& Partridge (1998). HP98 found that the star--forming galaxies that produce the detected 120--260 $\mu$m CIB intensity of $\sim$ 22 nW m$^{-2}$ sr$^{-1}$ contribute about 15~K to the 178~MHz  brightness temperature (they adopted a value of $\alpha$ = 0.7). This result seems at first glance to be in disagreement with the entry for the HP98 model in the table. The reason for this apparent "discrepancy" is that HP98 expressed the radio--IR correlation in terms of the FIR luminosity of galaxies. Had they expressed this correlation in terms of their IR luminosity, which is about twice the FIR value, they would have derived a radio brightness temperature of $\sim$ 17~K for a 3.5--1000~$\mu$m background intensity of 50 nW m$^{-2}$ sr$^{-1}$, almost identical to the value listed in the table. The analytical expression presented in this paper reproduces the results derived by HP98 for their specific CIB production scenario, and generalizes their treatment to any cosmic star formation history.

\section{DISCUSSION AND SUMMARY}

In this paper we derived an analytical expression for the correlation between the CIB and CRB intensities as a function of $\alpha$, the spectral index of the radio sources [see eq. (15)]. This correlation  is summarized for select galaxy number counts and CIB models in Table~1. Several conclusions can be drawn from a simple examination of the table: (1) the minimal CSFR,defined by the lower envelope of the data, is definitely ruled out as a viable representation of the CSFR since it falls short of providing the observed intensity of the CIB. The maximal CSFR,defined by the upper envelope of the data, is consistent with the CIB and CRB limits only for values of $\alpha \lesssim$ 0.8; (2) the baseline model of Malkan \& Stecker is barely consistent with the lower limit on the CIB intensity and consequently, for $\alpha$~=~0.7, it requires AGN to contribute more than 70\% of the 178~MHz background; (3) for $\alpha$~=~0.9, all models, with the exception of those of Malkan \& Stecker and Chary \& Elbaz, predict CRB temperatures that are 2$\sigma$ above the observed value. The latter two models leave no room for a significant contribution of AGN to the CRB; (4) for values of $\alpha \approx$ 0.6--0.8, the calculated $T_{\rm crb}/I_{\rm CIB}$ ratio is lower than the nominal observed ratios, suggesting that $\sim$60 to 20\% of the CRB at 178~MHz must arise from AGN, regardles of the radiative history of star--forming galaxies. 

All conclusions listed above assume that the radio--IR correlation observed in the local universe can extended to galaxies at all redshifts. Future observations will provide a larger sample of galaxies with which to study the radio--IR correlation and place tighter limits on the CIB, advances that will lead to a better understanding of the relative contribution of star--forming galaxies and AGN to the CRB.

Acknowledgement: We thank Rick Arendt for useful discussions and his insightful comments on an earlier version of the manuscript. ED acknowledges NASA's Astrophysics Theory Program NRA 99-OSS-01 for support of this work. MB's summer student internship at NASA/GSFC was supported by the "Research Opportunities for Undergraduates in the Laboratory for Astronomy and Solar Physics" program.

\clearpage

\begin{deluxetable}{llllll}
\tablecaption{CIB and CRB Intensities Predicted by Various Models 
for the Cosmic Star Formation Rate\tablenotemark{a}}
\tablehead{
\colhead{Model} &
\colhead{$I_{\rm CIB}$(nW m$^{-2}$ sr$^{-1}$)} &
  \multicolumn{4}{c}{$T_{\rm crb}$(K) at 178 MHz} \nl
% \colhead{$T_{CRB}$(K)\tablenotemark{c}} &
%& \colhead{$T_{CRB}$(K)}  \nl
\colhead{  } &
\colhead{ (3.5--1000 $\mu$m)} & 
 \colhead{ $\alpha$ = 0.6} &
 \colhead{ $\alpha$ = 0.7} &
 \colhead{ $\alpha$ = 0.8} &
 \colhead{$\alpha$ = 0.9} 
  }
\startdata 
minimum CSFR	                   &  8.5              & 2.7       & 3.1      &   3.5     &    4.1 \nl  
maximum CSFR                           &  73              & 24.8      & 28.0   &   31.5   &    35.6  \nl  
Haarsma \& Partridge (1998)	 & 50               & 15.7     & 18.1   &   20.8   &    24.1  \nl 
Malkan \& Stecker (2001)          & 23                & 8.2       & 9.1   &   10.2   &    11.4 \nl 
Chary \& Elbaz (2001)                & 37                & 12.3    & 13.9   &  15.8   &    18.0  \nl
Xu et al. (2001)                            & 58                & 18.8   & 21.5   &   24.5   &    27.9 \nl
Observational Limits\tablenotemark{b}       & 50$\pm$25  & 57$\pm$11 & 37$\pm$8 & 23$\pm$5 & 15$\pm$3  \nl
\hline
 ${\cal A}^{-1}\equiv T_{\rm crb}/I_{\rm CIB}$ (calculated)\tablenotemark{c} & &  0.33 & 0.37 & 0.42 & 0.48 \nl
$T_{\rm crb}/I_{\rm CIB}$ (observed)& &  1.1$\pm$0.6 & 0.7$\pm$0.4 & 0.5$\pm$0.25 & 0.30$\pm$0.16 \nl
                 
\enddata
\tablenotetext{a}{The CSFR predicted by the tabulated models are shown in Figure 2. CIB intensities and CRB temperatures were calculated from eqs. (10) and (11), respectively. The parameter $\alpha$ is the spectral index of the radio sources. }
\tablenotetext{b}{Observational limits on the CIB intensity are summarized in Hauser \& Dwek (2001). CRB temperatures were taken from Bridle (1967, Table VIII).}
\tablenotetext{c}{Calculated using eq. (15).}
\end{deluxetable}

%\clearpage

\end{document}